\documentclass[12pt,a4paper]{article}
\usepackage{graphicx}
\usepackage{amsmath}
\usepackage{amsfonts}
\usepackage{citesort}
\usepackage{amssymb}
\newcommand\wtktN%
{{\kern-3.8pt}\widetilde{{\kern3.3pt}N{\kern3.3pt}}{\kern-3.4pt}\rangle}
\newcommand\wtbrN%
{\langle{\kern-3.4pt}\widetilde{{\kern3.3pt}N{\kern3.3pt}}{\kern-3.4pt}}

\begin{document}
\hfill PTA/08-053
\bigskip

\begin{center}
{\Large{\bf Four-point function in Super Liouville Gravity}}
\end{center}
\vspace{1.0cm}
\begin{center}
{\large A.~Belavin}

\vspace{0.2cm}

L.~D.~Landau Institute for Theoretical Physics RAS

142432 Chernogolovka, Russia

\vspace{0.2cm}

and

\vspace{0.2cm}

{\large V.~Belavin}\footnote{Institute of
Theoretical and Experimental Physics, B.~Cheremushkinskaya 25, 117259 Moscow, Russia.}

\vspace{0.2cm}

Laboratoire de Physique Th\'eorique et Astroparticules

Universit\'e Montpellier II

Pl.~E.~Bataillon, 34095 Montpellier, France
\end{center}

\vspace{1.0cm}

\textbf{Abstract}

We consider the 2D super Liouville gravity coupled to the 
minimal superconformal theory. We analyze the physical states
in the theory and give the general form of the $n$-point correlation
numbers on the sphere in terms of integrals over the moduli space. 
The three-point correlation numbers are presented explicitly. For the 
four-point correlators, we show that the integral over the moduli space reduces
to the boundary terms if one of the fields is degenerate. It turns out
that special logarithmic fields are relevant for evaluating these boundary terms. We discuss the construction
 of these fields and study their operator product expansions.
This analysis allows evaluating the four-point correlation numbers. 
The derivation is analogous to the one in the bosonic case and is based on the recently derived higher equations 
of motion of the super Liouville field theory.

\section{Introduction}

Super Liouville gravity (SLG)~\cite{Polyakov1} is the two-dimensional
quantum gravity whose action is induced by a super conformal matter. This
induced action is universal (i.e., its form is independent of the concrete
choice of the conformal matter) and is known as the super Liouville action.
In the framework of the David and Distler--Kawai approach (DDK)~\cite{D,DK},
SLG is presented as a tensor product of a matter theory (SCFT), super
Liouville (SLFT) system, and super ghost (SG) system. The SG system also has
super conformal symmetry and appears as a result of the gauge-fixing problem
in SLG (see, e.g.,~\cite{DiFran} for details). Schematically, the SLG action
is
\begin{equation}
A_{\text{SLG}}=A_{\text{SCFT}}+A_{\text{SL}}+A_{\text{SG}}.
\end{equation}
All three theories constituting SLG are completely solvable by the standard
bootstrap technique~\cite{BPZ} (at least in principle) because of the
infinite symmetry given by the superconformal algebra
\begin{equation}
\begin{aligned}
\lbrack L_n,L_m]&=(n-m)L_{n+m}+\frac{c}8(n^3-n)\delta_{n;-m},
\\
\{G_r,G_s\}&=2L_{r+s}+\frac{c}2\left(r^2-\frac14\right)\delta_{r;-s},
\\
[L_n,G_r]&=\left(\frac12n-r\right)G_{n+r},
\end{aligned}
\label{The algebra}
\end{equation}
where $c$ is the central charge parameter and the 
indices $r$ and $s$ are integers for the Ramond (R) sector and half-integers for the Neveu--Schwarz (NS) sector.
We restrict ourself to considering only the NS sector here. Whenever
it cannot cause confusion, we omit sector indices for the superconformal
generators (i.e., indices indicating which sector---Liouville, matter, or
ghost---is under consideration).

The interaction between the three components of SLG in the DDK approach is
via the relation for the central charge parameters
\begin{equation}
c_{\text{SCFT}}+c_{\text{SL}}+c_{\text{SG}}=0
\label{totc}
\end{equation}
and also due to the construction of the physical fields and the integration
over the moduli space in constructing the correlation numbers (see below).

This paper is organized as follows. In the next three sections, we briefly
recall some relevant aspects concerning all three ingredients of SLG.
Section~5 is devoted to the preliminary analysis of the physical states in
SLG. We use the material in the preceding sections to construct the basic
types of the physical fields in the standard framework of BRST quantization.
We also introduce the special discrete series of physical states, known as 
``ground ring elements'' and discuss the relation between the 
logarithmic counterparts of the ground ring elements and the basic physical fields.
This is our first main result. In the next section, we discuss
the operator products of the ground ring elements with the basic physical
fields. In Sec.~7, we derive the general form of the $n$-point correlation
number on the sphere in SLG. This expression contains the integration over
the moduli space. The simplest case of three points, when the moduli space
is trivial, is given explicitly. For the four-point correlators, we show that
in special cases the integration over moduli reduces to the boundary terms, 
which are defined by the operator product expansions (OPEs) of the ``logarithmic''
fields. In Sec.~8, we calculate all necessary operator products explicitly
and evaluate the boundary terms. Finally, this consideration leads to an
explicit expression for the four-point correlation number of one degenerate
and three generic fields, which is presented in Sec.~9. This is our second
main result. Some calculations omitted in the main text are presented in
the appendices.

\section{Super Liouville field theory}

The super Liouville field theory is a super conformal field theory
with the central charge
\begin{equation}
c=1+2Q^{2},
\label{cQ}%
\end{equation}
where the ``background charge'' $Q$ is parameterized in terms of the basic
``quantum'' parameter $b$ as $Q=b^{-1}+b$. The classical (as well as the quantum) SLFT has been introduced 
and studied in \cite{Curtright,Arvis,DHoker,Babelon} shortly after it appeared in the string context
in \cite{Polyakov1}.  Below, we present the main facts
concerning NS sector of SLFT (see~\cite{BBNZ,VB1,LH1,LH2} and the references therein for
more details and recent developments). The NS fields belong to the 
highest-weight representations of the superconformal algebra. The basic fields are the scalar primary fields 
$V_{a}(x)$ corresponding to the highest-weight vectors
\begin{equation}
\begin{aligned}
L_{n}V_{a}&=0,\qquad\bar L_{n}V_{a}=0,\quad\text{for }n>0,
\\
G_{k}V_{a}&=0,\qquad\bar G_{k}V_{a}=0,\quad\text{for }k>0,
\\
L_{0}V_{a}&=\bar L_{0}V_{a}=\Delta_{a}V_{a},
\end{aligned}
\label{highest}
\end{equation}
where
\begin{equation}
\Delta_{a}=\frac{a(Q-a)}2
\label{Da}%
\end{equation}
and $a$ is a (complex) continuous parameter. We also use another parameter
$\lambda=Q/2-a$. The representations are singular at certain special
values of the parameters. This happens~\cite{Kac} at
$\lambda=\lambda_{m,n}$, where $(m,n)$ is a pair of positive integers
($m-n\in2\mathbb{Z}$) and
\begin{equation}
\lambda_{m,n}=\frac{mb^{-1}+nb}2.
\label{lmn}%
\end{equation}
In general, at $a=a_{m,n}$, one singular vector appears at the level $mn/2$
in the Verma module over $V_{a_{m,n}}=V_{m,n}$. For each pair $(m,n)$, it
is convenient to introduce a ``singular-vector creation operator'' $D_{m,n}$,
which is a graded polynomial in $G_{-k}$ and $L_{-k}$ of level $mn/2$ whose
coefficients are functions of the central charge parameter $b^{2}$ such that
the singular vector appears when $D_{m,n}$ is applied to $V_{m,n}$. The
normalization is unambiguously fixed via the coefficient of the
highest-order term $D_{m,n}=G_{-1/2}^{mn}+\dots$. The basic OPE is
\begin{equation}
V_{a_{1}}(x)V_{a_{2}}(0)=\int\frac{dP}{4\pi}
(x\bar x)^{\Delta-\Delta_{1}-\Delta_{2}}
(\mathbb{C}_{a_{1},a_{2}}^{Q/2+iP}[V_{Q/2+iP}(0)]_{\text{ee}}+
\tilde{\mathbb{C}}_{a_{1},a_{2}}^{Q/2+iP}
[V_{Q/2+iP}(0)]_{\text{oo}})
\label{VV}
\end{equation}
(for brevity here and hereafter, we set $\Delta=\Delta_{Q/2+iP}$ and
$\Delta_{i}=\Delta_{a_{i}}$). This OPE is continuous and involves
integration over the ``momentum'' $P$. In~(\ref{VV}), $[V_{p}]$ denotes
the contribution of the primary field $V_{p}$ and its superconformal
descendants to the OPE (subscript $\text{ee}$ stands for the descendants on the integer level and $\text{oo}$ -- for the
 descendants on the half-integer level). All other OPEs of two arbitrary
local fields
can be derived from~(\ref{VV}). The basic structure constants
$\mathbb{C}_{a_{1}a_{2}}^{Q/2+iP}$ and
$\tilde{\mathbb{C}}_{a_{1},a_{2}}^{Q/2+iP}$ in~(\ref{VV}) were
evaluated using the bootstrap technique in~\cite{Rubik,Marian,Fukuda}
and have the explicit form (here $a$ denotes $a_{1}+a_{2}+a_{3}$)
\begin{equation}
\begin{aligned}
\mathbb{C}_{a_{1}a_{2}}^{Q-a_3}&=\left(\!\pi\mu\gamma\!
\left(\frac{Qb}2\right)b^{1-b^2}\right)^{\!\!(Q-a)/b}\!
\frac{\Upsilon_{\text{R}}(b)\Upsilon_{\text{NS}}(2a_{1})
\Upsilon_{\text{NS}}(2a_{2})\Upsilon_{\text{NS}}(2a_{3})}
{2\Upsilon_{\text{NS}}(a-Q)\Upsilon_{\text{NS}}%
(a_{1+2-3})\Upsilon_{\text{NS}}(a_{2+3-1})\Upsilon_{\text{NS}}(a_{3+1-2})},
\\
\tilde{\mathbb{C}}_{a_{1}a_{2}}^{Q-a_3}&=-\left(\!\pi\mu\gamma\!
\left(\frac{Qb}2\right)b^{1-b^{2}}\right)^{\!\!(Q-a)/b}
\frac{i\Upsilon_{\text{R}}(b)\Upsilon_{\text{NS}}(2a_{1})
\Upsilon_{\text{NS}}(2a_{2})\Upsilon_{\text{NS}}(2a_{3})}
{\Upsilon_{\text{R}}(a-Q)\Upsilon_{\text{R}}%
(a_{1+2-3})\Upsilon_{\text{R}}(a_{2+3-1})\Upsilon_{\text{R}}(a_{3+1-2})},
\end{aligned}
\label{C3}
\end{equation}
where we use the convenient notation in~\cite{Fukuda}
for the special functions
\begin{equation}
\begin{aligned}
\Upsilon_{\text{NS}}(x)&=\Upsilon_{b}
\left(\frac x2\right)\Upsilon_{b}\left(\frac{x+Q}2\right),
\\
\Upsilon_{\text{R}}(x)&=\Upsilon_{b}
\left(\frac{x+b}2\right)\Upsilon_{b}\left(\frac{x+b^{-1}}2\right)
\end{aligned}
\label{YNSR}
\end{equation}
expressed in terms of the ``upsilon'' function $\Upsilon_{b}$, which is
standard in the Liouville field theory (see~\cite{DO, LFT}). Structure
constants~\eqref{C3} correspond to the normalization of the primary fields
\begin{equation}
\left\langle V_{a}V_{Q-a}\right\rangle _{\text{SLFT}}=
\left(x\bar x\right)  ^{-2\Delta_{a}}.
\label{Lnorm}%
\end{equation}
An important result concerning SLFT is the higher equations of motion.
Following~\cite{superhigher}, we define the set of ``logarithmic
degenerate fields'' in the NS sector
\begin{align}
V'_{m,n}&=\left.V'_{a}\right|_{a=a_{m,n}},\quad m-n\in2\mathbb{Z},
\label{VRprim}
\end{align}
where the general logarithmic fields $V'_{a}=\partial V_{a}/\partial a$
are the derivatives with respect to $a$ of the corresponding primary fields.
It turns out that while $V'_{m,n}$ are logarithmic fields (as well as general
$V'_{a}$), the fields
\begin{equation}
\bar D_{m,n}D_{m,n}V'_{m,n}
\label{DDV}%
\end{equation}
have the properties of primary fields and should be identified with the
exponential primary fields $V_{m,-n}$. More precisely, we have the relations
\begin{equation}
\bar D_{m,n}D_{m,n}V'_{m,n}=B_{m,n}V_{m,-n},
\label{HEM}%
\end{equation}
known as the higher equations of motion. Here, the exponential primaries
$V_{m,-n}$ have the dimensions $\Delta_{m,n}+mn/2$ and the coefficients
\begin{equation}
B_{m,n}=2^{mn}i^{mn-2[mn/2]}b^{n-m+1}[\pi\mu\gamma(bQ/2)]^{n}
\gamma\left(\frac{m-nb^{2}}2\right)\prod\nolimits_{(k,l)\in
\langle m,n\rangle_{\text{NS}}}\lambda_{k,l},
\label{oo}%
\end{equation}
where the set $\langle m,n\rangle_{\text{NS}}$ is the set of integer pairs
\begin{equation}
\bigl\{(k,l)\in\{k\mid1-m\le k\le m-1\},\{l\mid 1-n\le l\le n-1\}
\bigm|k-l\in2\mathbb{Z}\bigr\}\setminus\{(0,0)\}.
\label{oeset}
\end{equation}

\section{Generalized super minimal models}

We consider the special type of SLG where the so-called generalized
super minimal models (GSMM) are in the matter sector of the theory.
We call the corresponding induced Liouville gravity the minimal super
Liouville gravity. In the GSMM, there are no special restrictions on the
central charge, which can take an arbitrary value in principle, in contrast
to the case of ordinary minimal models. It is instructive to parameterize
the central charge via the same basic parameter $b$ as for SLFT,
\begin{equation}
c=1-2(b^{-1}-b)^{2}.
\label{cM}%
\end{equation}
In this parameterization, condition~\eqref{totc} for the total
central charge is satisfied automatically. The ``canonical'' super
minimal models appear for the special choice of the parameter $b$
such that $b^2$ is a rational number. Otherwise, the algebra of the
degenerate primary fields no longer closes within any finite subset;
instead, the whole set $\{\Phi _{m,n}\}$ with any pair $(m,n)$ of
natural numbers forms a closed algebra. Moreover, we enlarge the
space of local fields by including local fields with dimensions
different from the Kac values. Hence, the spectrum of dimensions
is continuous in GSMM. We introduce the continuous parameter $\alpha$
to parameterize a continuous family $\{\Phi_{\alpha}\}$ of primary
fields with the dimensions
\begin{equation}
\Delta_{\alpha}^{\text{(M)}}=\frac{\alpha(\alpha-q)}2,
\label{DM}%
\end{equation}
where
\begin{equation}
q=b^{-1}-b.
\label{q}%
\end{equation}
We also always use the ``canonical'' CFT normalization of the primary
fields $\Phi_{\alpha}$ via the two-point functions
\begin{equation}
\left\langle\Phi_{\alpha}\Phi_{\alpha}\right\rangle_{\text{GSMM}}=
\left(x\bar x\right)^{-2\Delta_{\alpha}}.
\label{Mnorm}%
\end{equation}
The degenerate fields $\Phi_{m,n}$ have the dimensions
\begin{equation}
\Delta_{m,n}^{\text{(M)}}=-q^{2}/8+\lambda_{m,-n}^{2}/2.
\label{DMatmn}%
\end{equation}
They correspond to either $\alpha=\alpha_{m,n}$ or $\alpha=q-\alpha_{m,n}$
with
\begin{equation}
\alpha_{m,n}=q/2+\lambda_{-m,n}.
\label{alphamn}%
\end{equation}
It can be seen that the construction of GSMM is formally similar to that
of SLFT and differs by the change $b\to ib$ and $\alpha\to-ia$
and also by the   normalization conditions for the primary fields,
Eq.~\eqref{Lnorm} for SLFT and Eq.~\eqref{Mnorm} for GSMM.

\section{Super ghosts}

In this section, we collect some results concerning SG (see,
e.g.,~\cite{Polchinski1, Verlinde,Fridan} for details). The SG appear as a
result of gauge fixing in the Polyakov approach to SLG and are described
by the free super conformal field theory with the central charge
$c_{\text{SG}}=-10$. 

The fermionic part of the SG system involves two
anticommuting fields $(b,c)$ of spins $(2,-1)$ with the action
\begin{equation}
A_{\text{bc}}=\frac{1}{2\pi}\int d^2z(b\bar\partial c+\bar b\partial\bar c).
\end{equation}
The operator products are readily found, with appropriate attention to the order
of the anticommuting variables,
\begin{equation}
b(z)c(0)\sim\frac1z,\qquad b(z)b(0)\sim O(z),\qquad c(z)c(0)\sim O(z).
\label{bcOPE}
\end{equation}
As usual we focus on the holomorphic part. The fields have the
Laurent expansions
\begin{equation}
b(z)=\sum_{m=-\infty}^{\infty}\frac{b_m}{z^{m+2}},\qquad
c(z)=\sum_{m=-\infty}^{\infty}\frac{c_m}{z^{m-1}},
\end{equation}
which yield the anticommutators
\begin{align}
\{b_m,c_n\}=\delta_{m+n,0},\qquad\{b_m,b_n\}=0,\qquad\{c_m,c_n\}=0.
\label{bcalgebra}
\end{align}
There are two natural ground states,
both are annihilated by $b_m$ and $c_m$ for $m>0$. In addition,
the first one is annihilated by $b_0$, and the second one is annihilated
by $c_0$. For future purposes, we choose the vacuum annihilated by $b_0$:
\begin{equation}
\begin{aligned}
b_m|v\rangle_{bc}&=0,\quad m\ge0,
\\
c_m|v\rangle_{bc}&=0,\quad m\ge1.
\end{aligned}
\end{equation}
It can be verified that the vacuum $|v\rangle_{bc}$ corresponds to the field $c(x)$
with the dimension $-1$. The unit operator, which is
relevant later, corresponds to the state
\begin{equation}
|1\rangle_{bc}=b_{-1}|v\rangle_{bc}.
\end{equation}
All other representations of~\eqref{bcalgebra} are equivalent to the one
described above.

The bosonic part of the SG involves two bosonic fields $(\beta,\gamma)$ of
spins $(3/2,-1/2)$ that are superpartners of the respective fermionic ghosts
$(b,c)$. The action has the form
\begin{equation}
A_{\beta\gamma}=\frac{1}{2\pi}\int d^2z
(\beta\bar\partial\gamma+\bar\beta\partial\bar\gamma).
\end{equation}
Because the statistics changes, some signs in the operator products change:
\begin{equation}
\beta(z)\gamma(0)\sim-\frac1z,\qquad\gamma(z)\beta(0)\sim\frac1z,\qquad
\beta(z)\beta(0)\sim O(1),\qquad\gamma(z)\gamma(0)\sim O(1).
\end{equation}
The Hilbert space of the $(\beta,\gamma)$-system is constructed by expanding
\begin{equation}
\beta(z)=\sum_{m=-\infty}^{\infty}\frac{\beta_m}{z^{m+3/2}},\qquad
\gamma(z)=\sum_{m=-\infty}^{\infty}\frac{\gamma_m}{z^{m-1/2}},
\end{equation}
where index $m$ is integer for the R sector and half-integer for the
NS sector and the coefficients satisfy the canonical commutation relations
\begin{align}
[\gamma_m,\beta_n]=\delta_{m+n,0},\qquad[\beta_m,\beta_n]=0,\qquad
[\gamma_m,\gamma_n]=0.
\label{betagammaalgebra}
\end{align}
The state-operator map is now trickier. In contrast to the fermionic case,
there now exist discreet series of vacuums $\{|q\rangle_{\beta\gamma}\}$
with $q$ either integer or half-integer,
\begin{equation}
\begin{aligned}
\beta_m|q\rangle_{\beta\gamma}&=0,\quad m\ge-q-1/2,
\\
\gamma_m|q\rangle_{\beta\gamma}&=0,\quad m\ge q+3/2.
\end{aligned}
\label{qvacuum}
\end{equation}
Each $q$-vacuum defines an inequivalent representation
of~\eqref{betagammaalgebra}, sometimes called a ``picture.'' States in
the Hilbert space are in a one-to-one correspondence with local field
operators, and the complete space of local fields is therefore the direct
sum of all highest-weight representations with the highest vectors
$|q\rangle_{\beta\gamma}$. We define the two local operators that are
most important for the subsequent developments. It can be verified that
the vacuum $|q=0\rangle_{\beta\gamma}$ has the dimension $0$ and
corresponds to the unit operator. The local operator corresponding to the
vacuum $|q=-1\rangle_{\beta\gamma}$ is particularly useful. It turns out
that this vacuum corresponds to the formal operator $\delta(\gamma(0))$
of dimension $1/2$. Indeed, taking the general properties of the Dirac
$\delta$-function and OPE~\eqref{betagammaalgebra} (see Appendix~A)
into account, we can straightforwardly derive the operator product
relations
\begin{equation}
\gamma(z)\delta(\gamma(0))\sim
z\partial\gamma(0)\delta(\gamma(0)),\qquad
\beta(z)\delta(\gamma(0))\sim-\frac 1z\delta(\gamma(0)),
\end{equation}
and similar equations with $\beta$ and $\gamma$ interchanged. Using
these expansions we can easily verify that $\delta(\gamma(0))$
satisfies~\eqref{qvacuum} for $q=-1$.

The anomaly in the conservation of the ghost currents $J^{bc}(u)=-{:}b(u)c(u){:}$ and
$J^{\beta\gamma}(u)=-{:}\beta(u)\gamma(u){:}$ leads to the
important requirement for the correlation functions. In particular,
on the sphere, any correlation function in SLG including ghost observables
built of $\beta$, $\gamma$, $\delta(\beta(0))$, $\delta(\gamma(0))$, and
their descendants should satisfy the ghost-number balance
\begin{equation}
\begin{aligned}
&N_c-N_b=3
\\
&(N_{\delta(\gamma)}-N_{\delta(\beta)})+(N_{\beta}-N_{\gamma})=2.
\end{aligned}
\label{gbalance}
\end{equation}
Although this relation is known (see, e.g.,~\cite{Polchinski1}), we
rederive this result in Appendix~A. Finally, the superconformal generators
in the ghost sector are
\begin{equation}
\begin{aligned}
L_m^g&=\sum_n(m+n)\,{:}b_{m-n}c_n{:}+\sum_k\biggl(\frac{m}{2}+k\biggr)\,
{:}\beta_{m-k}\gamma_k{:}-\frac{1}{2}\delta_{m,0},
\\
G_k^g&=-\sum_n \biggl[\biggl(k+\frac{n}{2}\biggr)\beta_{k-n}c_n+
2b_n\gamma_{k-n}\biggr],
\end{aligned}
\label{LGghost}
\end{equation}
where the normal ordering is defined with respect to the ground states
$|v\rangle_{bc}$ and $|q=-1\rangle_{\beta\gamma}$.
In the practical calculations, the computation relations of the ghost
generators to the superconformal generator $L_n$,
\begin{equation}
\begin{aligned}
&[L_n,b_m]=(n-m)b_{n+m},
\\
&[L_n,c_m]=-(2n+m)c_{n+m},
\\
&[L_n,\beta_k]=\biggl(\frac n2-k\biggr)\beta_{n+k},
\\
&[L_n,\gamma_k]=-\biggl(\frac{3n}2+k\biggr)\gamma_{n+k},
\end{aligned}
\end{equation}
and to the generator $G_k$,
\begin{equation}
\begin{aligned}
&[G_k,b_n]=\biggl(\frac n2-k\biggr)\beta_{k+n},
\\
&[G_k,c_n]=-2 \gamma_{k+n},
\\
&[G_k,\beta_r]=-2 b_{k+r},
\\
&[G_k,\gamma_r]=\biggl(\frac{3k}2 +\frac r2\biggr)c_{k+r},
\end{aligned}
\end{equation}
are relevant. In the next section, we use the facts described above to
construct the basic physical states in SLG.

\section{Physical states in SLG}

The vanishing of the total central charge~\eqref{totc} is not sufficient to ensure the Weyl invariance of SLG. 
Inserting the physical fields in the functional integral should also not
spoil Weyl invariance. This leads to a certain restriction on the
structure of the physical fields. The standard mathematical tool for
treating such problems is cohomology theory. The physical fields form a
space of cohomology classes with respect to the nilpotent BRST charge $Q$.
In modes, $Q$-operator can be derived via the commutation relations
\begin{equation}
\begin{aligned}
&\{b_n,Q\}=L_n,
\\
&[\beta_r,Q]=G_r,
\end{aligned}
\label{bQcomm}
\end{equation}
Here and below $L_n$ and $G_r$ are the total superconformal generators, defined by the sum of the corresponding
 generators in all three sectors (matter, Liouville,
and ghost),
 \begin{equation}
\begin{aligned}
&L_n=L_n^{\text{M}}+L_n^{\text{L}}+L_n^{\text{g}},
\\
&G_r=G_r^{\text{M}}+G_r^{\text{L}}+G_r^{\text{g}}.
\end{aligned}
\end{equation}
We also will use the notations $L_n^{\text{M+L}}=L_n^{\text{M}}+L_n^{\text{L}}$ and
$G_r^{\text{M+L}}=G_r^{\text{M}}+G_r^{\text{L}}$. The commutation relations~\eqref{bQcomm}
lead to the expansion of the BRST operator
\begin{align}
Q=\sum_m{:}\bigg[L_m^{\text{M+L}}+\frac{1}{2}L^{\text{g}}_m\bigg]c_{-m}{:}+
\sum_r{:}\bigg[G_r^{\text{M+L}}+\frac{1}{2}G^{\text{g}}_r\bigg]\gamma_{-r}{:}-\frac{1}{4}c_0,
\label{Q}
\end{align}
where $m$ ranges the integers, $r$ ranges the half-integers, and the normal
ordering is the same as in Eq.~\eqref{LGghost}. From this, we can verify
the nilpotence whenever the total central charge vanishes.
The observable spectrum consists of the BRST invariant local fields,
\begin{align}
Q|\Psi\rangle=0,
\label{QPhysSt}
\end{align}
which not belong to $\mathrm{Im}Q$.
Because of~\eqref{bQcomm}, the total dimension of any nontrivial physical state
is zero. Besides the local fields, for the construction of the physical amplitudes in SLG
we introduce also non-local physical fields of the form
\begin{align}
\int \bar b_{-1}b_{-1} \Psi(z,\bar z) d^2z,
\label{IntPhysSt}
\end{align}
where $\Psi(z,\bar z)$ satisfies~\eqref{QPhysSt}. The gauge invariance
of these fields follows directly from~\eqref{bQcomm}. We do not completely classify the physical states in SLG for
all different pictures here. Our aim here to bring sufficient information into
consideration to construct a gravitational $n$-point correlation number
for some special choice of the pictures.
As it was mentioned above, the ghost number balance~\eqref{gbalance} should be satisfied on the sphere. This condition
 can be fulfilled by choosing the fields in such a way that three of them have the ghost factors $c$ and $\bar c$ and two
 fields have the factors $\delta(\gamma)$ and $\delta(\bar\gamma)$. In particular it means that two of the field belong to
the picture $q=-1$ and others belong to the picture $q=0$.
We therefore concentrate on the physical states of this kind in what follows.
Apparently, the general case is obtained from this special one by applying the picture changing operator~\cite{Polchinski1}.

It is natural to start our search of the physical states among those
constructed from the primary fields in all three sectors. We call them the ``basic'' physical states. We introduce the
special notation for the vacuum vector in the tensor product of the matter
and Liouville sectors
\begin{equation}
|U_{a}\rangle=|\Phi_{a-b}\rangle|V_{a}\rangle.
\end{equation}
The vector $U_a$ corresponds to the field of total dimension
$(1/2,1/2)$. We also define the ground state related to the given
picture $q$:
\begin{equation}
|\Omega_a\rangle_q=|U_a\rangle|v\rangle_{bc}|q\rangle_{\beta\gamma}.
\end{equation}
It can be verified that
\begin{align}
|W_{a}\rangle=|\Omega_a\rangle_{-1}
\end{align}
satisfies~(\ref{QPhysSt}). Hence, the physical
fields of the first type are
\begin{align}
W_{a}(z,\bar z)=U_{a}(z,\bar z)\cdot c(z)\bar c(\bar z)\cdot
\delta(\gamma(z))\delta(\bar\gamma(\bar z)).
\label{W}
\end{align}
Similarly, it can be verified that another type of physical states can be
defined in the $q=0$ picture
\begin{align}
|\tilde W_{a}\rangle=\biggl(\bar G^{\text{M+L}}_{-1/2}+
\frac12\bar G_{-1/2}^{\text{g}}\biggr)\biggl(G^{\text{M+L}}_{-1/2}+
\frac12G_{-1/2}^{\text{g}}\biggr)|\Omega_a\rangle_{0}.
\end{align}
In the field language,
\begin{align}
{\tilde W}_{a}(z,\bar z)=\biggl(\bar G^{\text{M+L}}_{-1/2}+
\frac12\bar G_{-1/2}^{\text{g}}\biggr)\biggl(G^{\text{M+L}}_{-1/2}+
\frac12G_{-1/2}^{\text{g}}\biggr)U_{a}(z,\bar z)\cdot\bar c(\bar z)c(z).
\label{tildeW}
\end{align}
To give an idea of the explicit calculations, we prove that $\tilde W_a$
represents a cohomology class in Appendix~B.

The integral version of the physical state~\eqref{IntPhysSt},
corresponding to $\Psi(z,\bar z)=\tilde W_a (z,\bar z)$, is
\begin{align}
\int \bar G_{-1/2}G_{-1/2}U_{a}(z,\bar z)d^2z.
\label{intU}
\end{align}

An additional ``discrete'' physical state arises when the representation
in the matter sector is degenerate. The importance of the discrete
states~\cite{Discstates,Witten}, for calculating the correlators
in the bosonic LG was recently shown in~\cite{BZ1}. It is natural to assume
a similar effect in the supersymmetric extension of LG. We start by
describing the construction for the discrete states in SLG. The degenerate
matter fields $\Phi_{m,n}$, when combined with the degenerate exponentials
$V_{m,n}$ of the corresponding SLFT, yield nontrivial BRST invariant
operators of ghost number zero,
\begin{equation}
O_{m,n}(z,\bar z)=\bar H_{m,n}H_{m,n}\Phi_{m,n}(z,\bar z)V_{m,n}(z,\bar z).
\end{equation}
The operators $H_{m,n}$ are composed of the super Virasoro generators of the
level $(mn-1)/2$ in the matter and Liouville sectors and also of ghost
fields. The condition that the field $O_{m,n}$ is closed but nontrivial,
i.e., represents a cohomology class, defines the operator $H_{m,n}$
uniquely modulo exact terms. There is an important relation between the
discrete states and the physical operators discussed above. We introduce
the operator
\begin{equation}
J_{m,n}(z)=\biggl(G_{-1/2}^{\text{M+L}}-\frac12G_{-1/2}^{\text{g}}\biggr)D_{m,n}^{\text{L}} c(z),
\label{Jmn}
\end{equation}
where the action of this operator is simply defined by the left
multiplication. Below we omit the dependence on $z$ and $\bar z$. Then the basic relation
\begin{equation}
\bar QQO_{m,n}=\bar{J}_{m,n}J_{m,n}\Phi_{m,n}V_{m,n}
\label{basic0}
\end{equation}
holds, where we temporarily assume that the singular vector in the Liouville sector
is not decoupled. In Appendix~C, we prove this relation and also use it to
derive the important statement
\begin{equation}
 \bar QQO'_{m,n}=B_{m,n}\tilde W_{m,-n},
\label{basic}
\end{equation}
where we introduce the logarithmic counterparts of the discrete states
$O_{m,n}$,
\begin{equation}
O'_{m,n}=\bar H_{m,n}H_{m,n}\Phi_{m,n}V_{m,n}',
\end{equation}
and $B_{m,n}$ is just the coefficients in the higher equations of motion
of SLFT~\eqref{HEM}. We note that in the space of states enlarged by
the logarithmic fields, the field $\tilde W_{m,n}$ becomes
trivial.\footnote{Naively, this would lead to trivial results for the
correlation functions with any other physical fields. But we see below
that because of the transformation properties of the BRST current, this
does not happen.}

Another useful consequence of~\eqref{basic} can be obtained by applying
$\bar b_{-1}b_{-1}$,
\begin{align}
\bar G_{-1/2}G_{-1/2}U_{m,-n}=B_{m,n}^{-1}
\left(\bar\partial-\bar Q \bar b_{-1}\right)
\left(\partial-Qb_{-1}\right)O_{m,n}'
\label{GGU}
\end{align}
or
\begin{align}
\bar G_{-1/2}G_{-1/2}U_{m,-n}=
B_{m,n}^{-1}\bar\partial\partial O'_{m,n}\mod Q.
\label{GGU1}
\end{align}
This means that the integrand in~\eqref{intU} is the full derivative
modulo BRST exact terms.

Apparently~\cite{Feigin}, the particular cases described above correspond
to the following general structure of the space of physical states in SLG.
In each picture, there is only one cohomology if the representation in
the matter sector is not degenerate. In the degenerate case, two additional physical fields arise. For example,
in the picture $q=0$, there are fields
$\tilde W_{m,n}$ and $O_{m,n}$ and the third type of cohomology of the ghost
number $N_g=2$, which we do not discuss here.

\section{Ground ring operator products}

The discrete states $O_{m,n}$ act modulo exact forms in the space of physical
states of the given type (in our case either $W_a$ or $\tilde W_a$) because their action does not change the ghost number
and all nontrivial classes in the given pictures
are generically exhausted by these composite fields with different $a$.
Moreover, because of the fusion restrictions of the degenerate fields
$\Phi_{m,n}$ and $V_{m,n}$ in the OPE, the general structure of the
operator products is
\begin{equation}
\begin{aligned}
O_{m,n}W(a)&=\sum_{\{r,s\}\in(m,n)}A_{r,s}^{(m,n)}W(a+\lambda_{r,s}),
\\
O_{m,n}\tilde W(a)&=\sum_{\{r,s\}\in(m,n)}\tilde A_{r,s}^{(m,n)}
\tilde W(a+\lambda_{r,s}),
\end{aligned}
\label{OmnW}
\end{equation}
where the set of integers $(m,n)$ is defined by the restrictions on the
OPE of the degenerate fields in both the matter and the Liouville sectors.
To evaluate the numerical coefficients $A_{r,s}^{(m,n)}$ and
$\tilde A_{r,s}^{(m,n)}$, it is useful to calculate explicitly in the
simplest nontrivial case $(m,n)=(1,3)$. In this case, $a_{1,3}=-b$.
Equation~\eqref{basic0} defines $H_{mn}$ explicitly. For $H_{1,3}$, we find
\begin{align}
H_{1,3}=L^{\text{M}}_{-1}-L^{\text{L}}_{-1}-G^{\text{M}}_{-1/2}G^{\text{L}}_{-1/2}+b^2\beta_{-3/2}\gamma_{1/2}+
2b^2b_{-2}c_1-b^2(G^{\text{M}}_{-1/2}+G^{\text{L}}_{-1/2})\beta_{-3/2}c_1
\end{align}
and the corresponding field
\begin{align}
O_{13}(x)={}&\Phi_{13}'(x)V_{13}(x)-\Phi_{13}(x)V_{13}'(x)-
\Psi_{13}(x)\Lambda_{13}(x)+[b^2\,{:}\beta(x)\gamma(x){:}
\nonumber
\\
&{}+2b^2\,{:}b(x)c(x){:}]\Phi_{13}(x)V_{13}(x)-
b^2\beta(x)c(x)\Psi_{13}(x)V_{13}(z)
\nonumber
\\
&{}-b^2\beta(x)c(x)\Phi_{13}(x)\Lambda_{13}(z).
\label{O13}
\end{align}
Here, we introduce the special notation for the top components of the
primary supermultiplets in the Liouville\footnote{In the  Liouville sector,
our notation for the top component of the primary supermultiplet differs
from the notation in~\cite{BBNZ,VB1}.} and matter sectors
\begin{equation}
\begin{aligned}
&\Lambda_a=\bar G^{\text{L}}_{-1/2}G^{\text{L}}_{1/2}V_a,
\\
&\Psi_a=\bar G^{\text{M}}_{-1/2} G^{\text{M}}_{1/2}\Phi_a.
\end{aligned}
\end{equation}
We first consider the operator product $O_{13}(x)W_a (0)$. The special
OPE that we need in this case are those of the degenerate field  $V_{13}$
and $\Lambda_{13}$ in the Liouville sector (see~\cite{BBNZ,VB1} for details)
\begin{equation}
\begin{aligned}
&V_{1,3}(x)V_{a}(0)=(x\bar x)^{ab}C_{+}^{\text{L}}(a)[V_{a-b}]_{\text{ee}}
\\
&\phantom{V_{1,3}(x)V_{a}(0)={}}+
(x\bar x)^{1+b^{2}}\tilde C_{0}^{\text{L}}(a)[V_{a}]_{\text{oo}}+
(x\bar x)^{1-ba+b^{2}}C_{-}^{\text{L}}(a)[V_{a+b}]_{\text{ee}},
\\
&\Lambda_{1,3}(x)V_{a}(0)=(x\bar x)^{ab+1/2}C_{+}^{\text{L}}(a)
[V_{a-b}]_{\text{oo}}
\\
&\phantom{\Lambda_{1,3}(x)V_{a}(0)={}}+(x\bar x)^{b^{2}}
\tilde C_{0}^{\text{L}}(a)[V_{a}]_{\text{ee}}+(x\bar x)^{1-ba+b^{2}}
C_{-}^{\text{L}}(a)[V_{a+b}]_{\text{oo}},
\end{aligned}
\label{OPE13}
\end{equation}
where $C_{-}(a)$, $\tilde C_{0}(a)$, and $C_{+}(a)$ are ``special''
structure constants\footnote{Again, here we use the notation for the
Liouville special structure constants differently (which seems more
natural) than in~\cite{BBNZ,VB1}.}
\begin{equation}
\begin{aligned}
&C_{+}^{\text{L}}(a)=1,
\\
&\tilde C_{0}^{\text{L}}(a)=\frac{2\pi i\mu\gamma(ab-b^{2})}
{\gamma(-b^{2})\gamma(ab)},
\\
&C_{-}^{\text{L}}(a)=\pi^{2}\mu^{2}b^{4}\gamma^{2}
\left(\frac12+\frac{b^{2}}2\right)
\gamma\left(-\frac12-\frac{b^{2}}2+ab\right)
\gamma\left(\frac12-\frac{b^{2}}2-ab\right).
\end{aligned}
\label{C+-0Liouv}
\end{equation}
The analogous operator products in the matter sector can be easily
reconstructed by starting from~\eqref{OPE13}, renormalizing the
fields, and analytically continuing, as discussed in Sec.~3,
\begin{equation}
\begin{aligned}
&\Phi_{13}(x)\Phi_{a}(0)=(x\bar x)^{1-a b-b^2}C_{-}^{\text{M}}(a)
[\Phi_{a-b}]_{\text{ee}}
\\
&\phantom{\Phi_{13}(x)\Phi_{a}(0)={}}+(x\bar x)^{1-b^2}
\tilde C_{0}^{\text{M}}(a)[\Phi_{a}]_{\text{oo}}+
(x\bar x)^{a b}C_{+}^{\text{M}}(a)[\Phi_{a+b}]_{\text{ee}},
\\
&\Psi_{13}(x)\Phi_{a}(0)=(x\bar x)^{1-a b-b^2}C_{-}^{\text{M}}(a)
[\Phi_{a-b}]_{\text{oo}}
\\
&\phantom{\Psi_{13}(x)\Phi_{a}(0)={}}+(x\bar x)^{-b^2}
\tilde C_{0}^{\text{M}}(a)[\Phi_{a}]_{\text{ee}}+
(x\bar x)^{a b}C_{+}^{\text{M}}(a)[\Phi_{a+b}]_{\text{oo}}
\end{aligned}
\end{equation}
and
\begin{equation}
\begin{aligned}
&C_{-}^{\text{M}}(a)=\biggl(\frac{\gamma(1/2+b^{2}/2)\gamma(ab-1/2+b^{2}/2)}
{\gamma(-1/2+3b^{2}/2)\gamma(ab+1/2-b^{2}/2)}\biggr)^{1/2},
\\
&\tilde C_{0}^{\text{M}}(a)=ib^{-2}\gamma\left(\frac{bQ}{2}\right)
\biggl(\frac{\gamma(1-b^{2})\gamma(b^{2}/2-1/2)}
{\gamma(b^{2}-1)\gamma(3b^{2}/2-1/2)}\biggr)^{1/2}
\frac{\gamma(ab+b^{2})}{\gamma(ab)},
\\
&C_{+}^{\text{M}}(a)=\biggl(\frac{\gamma(1/2+b^{2}/2)
\gamma(ab-1/2+3b^{2}/2)}{\gamma(-1/2+3b^{2}/2)
\gamma(ab+1/2+b^{2}/2)}\biggr)^{1/2}.
\end{aligned}
\end{equation}
It is straightforward to verify that we are left with
\begin{equation}
O_{13}(x)W_a(0)=A_{0,-2}^{(1,3)}W_{a-b}(0)+
A_{0,0}^{(1,3)}W_{a}(0)+A_{0,2}^{(1,3)}W_{a+b}(0)
\label{O13W}
\end{equation}
in the operator product (see Appendix~A). The cancellation of the
``unphysical'' terms in the operator product can be verified by
explicitly calculating at least at the primary field level. The
coefficients can be written in the special factored form
\begin{equation}
\begin{aligned}
&A_{0,-2}^{(1,3)}=(1-2ab+b^2)^2C_{-}^{\text{M}}(a-b)
C_{+}^{\text{L}}(a)=X\frac{N(a)}{N(a-b)},
\\
&A_{0,0}^{(1,3)}=\tilde C_{0}^{\text{M}}(a-b)
\tilde C_{0}^{\text{L}}(a)=X\frac{N(a)}{N(a)},
\\
&A_{0,2}^{(1,3)}=(1-2 a b+b^2)^2C_{+}^{\text{M}}(a-b)
C_{-}^{\text{L}}(a)=X\frac{N(a)}{N(a+b)},
\end{aligned}
\end{equation}
where
\begin{equation}
X=4b^2\biggl[\pi\mu\gamma\biggl(\frac12+\frac{b^2}2\biggr)\biggr]
\biggl[\frac{\gamma(1/2+b^2/2)}{\gamma(3b^2/2-1/2)}\biggr]^{1/2}
\end{equation}
and
\begin{equation}
N(a)=\biggl[\pi\mu\gamma\biggl(\frac12+\frac{b^2}2\biggr)\biggr]^{-a/b}
\biggl[\gamma\biggl(ab-\frac{b^2}2+\frac12\biggr)
\gamma\biggl(\frac ab-\frac{b^{-2}}2+\frac12\biggr)\biggr]^{1/2}.
\label{N}
\end{equation}
A similar calculation can be performed for the operator product $O_{m,n}\tilde W$. 
It turns out that the result can be generalized for an arbitrary pair $(m,n)$ as
\begin{equation}
A_{r,s}^{(m,n)}=\tilde A_{r,s}^{(m,n)}=
K B_{m,n}N(a_{m,-n})\frac{N(a)}{N(a+\lambda_{r,s})},
\end{equation}
and the coefficient $K$ is universal, i.e., is independent of $(m,n)$,
\begin{equation}
K=\frac{1}{2 b}
\biggl[\frac{\gamma(1/2+b^2/2)}{\gamma(3/2-b^{-2}/2)}\biggr]^{1/2}.
\end{equation}
Expression~\eqref{OmnW} reduces to
\begin{equation}
\begin{aligned}
&O_{m,n}\frac{W(a)}{N(a)}= K B_{m,n}N(a_{m,-n})\sum_{\{r,s\}\in(m,n)}
\frac{W(a+\lambda_{r,s})}{N(a+\lambda_{r,s})},
\\
&O_{m,n}\frac{\tilde W(a)}{N(a)}=K B_{m,n}N(a_{m,-n})\sum_{\{r,s\}\in(m,n)}
\frac{\tilde W(a+\lambda_{r,s})}{N(a+\lambda_{r,s})}.
\end{aligned}
\label{OmnW1}
\end{equation}

\section{Correlation numbers}

The requirement that the ghost current be conserved leads to the general
form of the $n$-point correlation numbers on the sphere for the three basic types of
observables introduced in the previous section:
\begin{equation}
\langle\langle a_1a_2\cdots a_n\rangle\rangle_{\text{SLG}}=
\biggl\langle W_{a_{1}}(z_{1})\,W_{a_{2}}(z_{2})\,\tilde W_{a_{3}}(z_{3})
\prod_{i=4}^{n}\int\bar G_{-1/2}G_{-1/2}U_{a_i}(z_i)d^2z_i\biggr\rangle.
\label{corrSMG}
\end{equation}
The simplest case is the three-point correlation number, where there is
no integral over the moduli space and the result is factored into a
product of the matter, Liouville, and ghost three-point functions.
Using~\eqref{W} and~\eqref{tildeW}, we can derive
\begin{align}
\langle\langle a_1a_2a_3\rangle\rangle_{\text{SLG}}={}&
[\langle V_{a_1}V_{a_2}V_{a_3}\rangle
\langle\Phi_{a_1-b}\Phi_{a_2-b}\Psi_{a_3-b}\rangle
\nonumber
\\
&\;{}+\langle V_{a_1}V_{a_2}\Lambda_{a_3}\rangle
\langle\Phi_{a_1-b}\Phi_{a_2-b}\Phi_{a_3-b}\rangle]
\langle c_1c_2c_3\rangle\langle\delta(\gamma_1)\delta(\gamma_2)\rangle
\nonumber
\\
={}&C^{\text{M}}(a_1-b,a_2-b,a_3-b)\tilde C^{\text{L}}(a_1,a_2,a_3)
\nonumber
\\
&{}+\tilde C^{\text{M}}(a_1-b,a_2-b,a_3-b)C^{\text{L}}(a_1,a_2,a_3)
\label{3point}
\end{align}
or explicitly (see Appendix~E)
\begin{equation}
\langle\langle a_1a_2a_3\rangle\rangle_{\text{SLG}}=\Omega\prod_{i=1}^3N(a_i),
\label{3point1}
\end{equation}
where
\begin{equation}
\Omega=i\biggl[\pi\mu\gamma\biggl(\frac12+\frac{b^2}2\biggr)\biggr]^{Q/b}
\biggl[\frac{\gamma(b^2/2+1/2)\gamma(b^{-2}/2-1/2)}{b^2}\biggr]^{1/2}
\label{Omega}
\end{equation}
and the normalization factor $N(a)$ is defined in~\eqref{N}.

The partition sum and the two-point numbers can be obtained simply from the
expression for the three-point function. The next step is to obtain the
four-point numbers. The expression is more complicated: it involves
integration over moduli. The naive and rather numerical way to calculate
it is based on the conformal block decomposition of the four-point
correlation functions in both the Liouville and the matter sectors (see,~\cite{BBNZ,LH3,VB0}). 
But this approach is not available at the moment (we
intend to investigate it later), and this straightforward computation is
moreover unable to provide the exact results. Here, we use another
approach to evaluate the four-point integral. Relation~\eqref{GGU1} allows
reducing the moduli integral in expression~\eqref{corrSMG} for the
correlation numbers to the boundary integrals if one of the parameters is 
degenerate, $a_i=a_{m,-n}$. In
particular, the four-point correlation number becomes
\begin{equation}
\langle\langle a_{m,-n}a_1a_2a_3\rangle\rangle_{\text{SLG}}=
B_{m,n}^{-1}\int_{\partial\Gamma}\partial \langle
O_{m,n}'(x)W_{a_{1}}(x_{1})\, W_{a_{2}}(x_{2})\,
\tilde W_{a_{3}}(x_{3})\rangle\frac{dx}{2i},
\label{corrSMG1}
\end{equation}
where the boundary consists of three circles $\partial\Gamma=
\sum_{i=1}^3\partial\Gamma_i$ around the points $x_1$, $x_2$, and
$x_3$ (integrated clockwise) and a large circle
$\partial\Gamma_{\infty}$ near infinity (integrated counterclockwise), 
which arises because
the operator $O'_{m,n}$ is not exactly a scalar. Taking the distributive
character of the action of the BRST charge into account, we can shift it
after the insertion of the other physical fields, which means that
$Q$-exact terms in~\eqref{GGU1} do not contribute to~\eqref{corrSMG1}.
To evaluate the boundary terms, we must better understand the short-range
behavior of the operator products $O'_{m,n}(x)W(0)$ and
$O'_{m,n}(x)\tilde W(0)$. The next section is devoted to this subject.

\section{Boundary terms}

The boundary integrals in~\eqref{corrSMG1} are controlled by the operator products of
the logarithmic fields $O_{m,n}'$ with the basic physical states $W(a_i)$
and $\tilde W(a_i)$. The derivation of these OPEs on the basis of the corresponding ground ring operator products is 
almost the same as in the
bosonic case~\cite{BZ1,BZ11}. The only differences are the different definition
of the logarithmic field $V'(a)$ and the different relation between
the conformal dimension and the parameter $b$. A closer consideration
shows that these two differences compensate each other. Hence, the necessary logarithmic 
contributions to the OPE are
\begin{equation}
\begin{aligned}
&O'_{m,n}\frac{W(a)}{N(a)}=\log(x\bar x)K B_{m,n}N(a_{m,-n})
\sum_{\{r,s\}\in(m,n)}q_{r,s}^{(m,n)}(a)
\frac{W(a+\lambda_{r,s})}{N(a+\lambda_{r,s})},
\\
&O'_{m,n}\frac{\tilde W(a)}{N(a)}=\log(x\bar x)K B_{m,n}N(a_{m,-n})
\sum_{\{r,s\}\in(m,n)}q_{r,s}^{(m,n)}(a)
\frac{\tilde W(a+\lambda_{r,s})}{N(a+\lambda_{r,s})},
\end{aligned}
\label{O'mnW}
\end{equation}
where
\begin{equation}
q_{r,s}^{(m,n)}(a)=|a-\lambda_{r,s}-Q/2|_{\text{Re}}-\lambda_{m,n}
\end{equation}
and
\begin{equation}
|x|_{\text{Re}}=
\genfrac{\{}{.}{0pt}{}{\;\;x\quad\text{if }\operatorname*{Re}x>0,}
{-x\quad\text{if }\operatorname*{Re}x<0.}
\label{are}%
\end{equation}
The contribution at infinity arises as follows. The logarithmic field is
not a scalar; under conformal coordinate transformations $x\to y$, it
acquires an inhomogeneous part
\begin{equation}
O'_{m,n}(y)=O'_{m,n}(x)-2\Delta'_{m,n}O_{m,n}(x)\log|y_x|,
\label{Olog}%
\end{equation}
where
\begin{equation}
\Delta'_{m,n}=
\frac d{da}\Delta_{a}^{\text{(L)}}|_{a=a_{m,n}}=\lambda_{m,n}.
\label{Dmnp}%
\end{equation}
Transformation~(\ref{Olog}) leads to the behavior of the correlation
function with $O'_{m,n}(x)$ as $x\to\infty$%
\begin{equation}
\langle O'_{m,n}(x)W_{a_1}(x_1)W_{a_2}(x_2)\tilde W_{a_3}(x_3)\rangle\sim
-2\Delta'_{m,n}\log(x\bar x)\langle O_{m,n}W_{a_1}W_{a_2}\tilde W_{a_3}\rangle.
\label{Opinf}%
\end{equation}
Therefore, the contribution of the boundary term $\partial\Gamma_{\infty}$
is evaluated as
\begin{equation}
\frac1{2i}\int_{\partial\Gamma_{\infty}}\partial
\langle O'_{m,n}(x)W_{a_1}(x_1)W_{a_2}(x_2)\tilde W_{a_3}(x_3)\rangle dx=
-2\pi\lambda_{m,n}\langle O_{m,n}W_{a_1}W_{a_2}\tilde W_{a_3}\rangle.
\label{G4curv}%
\end{equation}

\section{Four-point correlation number}

Summing boundary contributions~(\ref{O'mnW}) and curvature
term~(\ref{G4curv}), we find the expression for the four-point
correlation number
\begin{equation}
\langle\langle a_{m,-n}a_1a_2a_3\rangle\rangle_{\text{SLG}}=\pi
K N(a_{m,-n})\biggl\{\sum_{i=1}^3\sum_{r,s\in (m,n)}
q_{r,s}^{(m,n)}(a_i)+2 m n\lambda_{m,n}\biggr\}
\langle\langle a_1a_2a_3\rangle\rangle,
\label{4point}
\end{equation}
where the fusion set $(m,n)=\{1-m:2:m-1,1-n:2: n-1\}$
and the second term is just the result of the curvature
contribution when we move $O_{m,n}'$ close to one of the other fields.
The normalization factor $N(a_{m,-n})$ is defined by~\eqref{N}, and the three-point
correlation number $\langle\langle a_1a_2a_3\rangle\rangle$ is given
by~\eqref{3point1}. It seems tempting to simplify these relations by
introducing the renormalized fields $\mathcal{W}_a$, $\mathcal{\tilde W}_a$,
and $\mathcal{U}_a$ as
\begin{equation}
\begin{aligned}
&\mathcal{W}_a=\frac{W_a}{N(a)},
\\
&\mathcal{\tilde W}_a=\frac{\tilde W_a}{N(a)},
\\
&\mathcal{U}_{a}=\frac{U_a}{N(a)}.
\end{aligned}
\end{equation}
Expression (\ref{4point}) reduces to
\begin{align}
\frac1{\Omega}\int\langle\bar G_{-1/2}G_{-1/2}\mathcal{U}_{m,-n}(x)
\mathcal{W}_{a_1}(x_1)\mathcal{W}_{a_2}(x_2)
\mathcal{\tilde W}_{a_3}(x_3)\rangle d^2 x
\nonumber
\\
=\pi K \biggl\{\sum_{i=1}^3
\sum_{r,s\in (m,n)}q_{r,s}^{(m,n)}(a_i)+2 m n\lambda_{m,n}\biggr\},
\label{4point1}
\end{align}
where $\Omega$ is defined in~\eqref{Omega}. The formulas~\eqref{4point} and~\eqref{4point1}
for the four-point correlation numbers in the minimal super Liouville gravity, 
together with the explicit expression for the structure constants~\eqref{3point1}, are the main results of the 
presented study.

\medskip

\textbf{Acknowledgments}
Authors are grateful to M.~Bershtein, D.~Fri\-ed\-an, A.~Litvinov, A.~Losev and 
D.~Po\-lya\-kov for the useful discussions. A.~B.~was supported by the RAS 
program ``Elementary Particles and Fundamental Nuclear Physics'' 
and the Russian Foundation for
Basic Research (Grant No.~07-02-00799) and also by Grant No.~SS-3472.2008.2.
V.~B.~was supported by  the Russian Foundation for Basic Research (Grant
No.~08-01-00720). V.~B. sincerely thanks A.~Neveu and other members
of LPTA, Montpellier University II, for their hospitality and the interest 
in this work.

\medskip

\appendix{\bf Appendix A. Ghost number balance on the sphere.}

Conservation of the fermionic ghost current provides the relation
\begin{equation}
\oint_C\frac{d u}{2\pi i}\langle J^{bc}(u)X(z_1,\dots,z_n)\rangle=
(N_c-N_b)\langle X(z_1,\dots,z_n)\rangle,
\label{intJX}
\end{equation}
where $X(z_1,\dots,z_n)$ denotes a set of the physical fields, the contour
$C$ is any contour encircling all insertions of the fields, and $N_c$
and $N_b$ are the total ghost numbers related to the composite operator
$X$. On the other hand, we can deform the contour, moving it to infinity.
For this, we need to know the transformation law for $J^{bc}$. First, we
define the infinitesimal version. Based on the canonical operator
products
\begin{equation}
\begin{aligned}
T(u)b(z)=\frac2{(u-z)^2}b(z)+\frac1{u-z}b'(z),
\\
T(u)c(z)=\frac{-1}{(u-z)^2}c(z)+\frac1{u-z}c'(z),
\end{aligned}
\end{equation}
correspond to the following ghost part of the stress-energy tensor 
\begin{equation}
T^{bc}(u)={:}c(u)b(u){:}+2\,{:}\partial c(u)b(u){:}\,,
\end{equation}
which defines (see~\eqref{bcOPE}) the singular part of the operator product
\begin{equation}
T^{bc}(u)J^{bc}(z)=-\frac3{(u-z)^3}+\frac{J^{bc}(z)}{(u-z)^2}+
\frac{\partial J^{bc}(z)}{(u-z)}.
\end{equation}
This, together with the definition
\begin{equation}
\delta_{\epsilon}=\oint\epsilon(u)T(u)\frac{du}{2\pi i},
\end{equation}
leads to
\begin{equation}
\delta_{\epsilon}J^{bc}(z)=-\frac32\epsilon''(z)+\epsilon'(z)J^{bc}(z)+
\epsilon\frac{\partial J^{bc}(z)}{\partial z}.
\label{inftes}
\end{equation}
This infinitesimal form allows reconstructing the finite version of the
transformation
\begin{equation}
J^{bc}(z)\to\tilde J^{bc}(z)=\frac{dw}{dz}J^{bc}(w(z))-
\frac32\frac{w''}{w'}\quad\text{as }z\to w(z).
\end{equation}
Also taking the transformation properties of the physical fields into
account, we now make the inversion for the correlation function
in~\eqref{intJX},
\begin{align}
\oint_C \frac{du}{2\pi i}\langle J^{bc}(u)X \rangle&=
\oint_{C_{\infty}}\frac{du}{2\pi i}[-\frac1{u^2}
\langle J^{bc}(1/u)X(1/z_1,\dots,1/z_n)\rangle+
\frac3u\langle X(1/z_1,\dots,1/z_n)\rangle]
\nonumber
\\
&=3\langle X(1/z_1,\dots,1/z_n)\rangle,
\end{align}
where the first term vanishes because $1/u=0$ is the regular point of the
correlation function. Again performing the inversion for the correlator
$\langle X(1/z_1,\dots,1/z_n)\rangle$, we obtain
\begin{equation}
(N_c-N_b-3)\langle X(z_1,\dots,z_n)\rangle=0,
\end{equation}
which means that either $N_c-N_b=3$ or the corresponding correlator is
equal to zero.

The consideration for the bosonic ghost current $J^{\beta\gamma}(z)$
literally follows that for the fermionic ghost current. We consider the
integral
\begin{align}
&\oint_C\frac{du}{2\pi i}\langle J^{\beta\gamma}(u)X(z_1,\dots,z_n)\rangle=
(-N_{\delta(\gamma)}+N_{\gamma}+N_{\delta(\beta)}-N_{\beta})
\langle X(z_1,\dots,z_n) \rangle,
\label{intJX2}
\\
&T^{\beta\gamma}(u)=-\frac12\,{:}\beta'(u)\gamma(u){:}-
\frac32\,{:}\gamma'(u)\beta(u){:}\,,
\end{align}
and
\begin{equation}
T^{\beta\gamma}(u)J^{\beta\gamma}(z)=
\frac2{(u-z)^3}+\frac{J^{\beta\gamma}(z)}{(u-z)^2}+
\frac{{\partial J^{\beta\gamma}}(z)}{(u-z)}.
\end{equation}
Hence, the infinitesimal form of the transformation is
\begin{equation}
\delta_{\epsilon}J^{\beta\gamma}(z)=
\epsilon''(z)+\epsilon'(z) J^{\beta\gamma}(z)+
\epsilon \frac{\partial J^{\beta\gamma}(z)}{\partial z},
\label{inftess}
\end{equation}
and the finite version is
\begin{align}
J^{\beta\gamma}(z)\to\tilde J^{\beta\gamma}(z)=
\frac{d w}{d z} J^{\beta\gamma}(w(z))+\frac{w''}{w'}\quad
\text{as }z\to w(z).
\end{align}
The only difference comes from the different coefficient in the
transformation law. We conclude that
\begin{equation}
(-N_{\delta(\gamma)}+N_{\gamma}+N_{\delta(\beta)}-N_{\beta}+2)
\langle X(z_1,\dots,z_n)\rangle=0
\end{equation}

\medskip

\appendix{\bf Appendix B. BRST properties of the field ${\tilde W}_{a}$.}

Here, we prove that ${\tilde W}_{a}$ is closed. We must verify that
\begin{equation}
Q {\tilde W}_{a}=Q(G^{\text{M+L}}_{-1/2}+
\frac12G_{-1/2}^g)|\Omega_a\rangle_0=
Q(G^{\text{M+L}}_{-1/2}-b_{-1}\gamma_{1/2})|\Omega_a\rangle_0=0,
\end{equation}
where the second equality follows from the mode expansion of $G_{1/2}^g$.
In accordance with definition~\eqref{Q}, we split operator $Q$ into three
parts $Q_1$, $Q_2$, and $Q_3$. Then
\begin{equation}
\begin{aligned}
&Q_1G^{\text{M+L}}_{-1/2}|\Omega_a\rangle_0=\frac12c_0G_{-1/2}|\Omega_a\rangle_0,
\\
&Q_2G^{\text{M+L}}_{-1/2}|\Omega_a\rangle_0=\bigg(\gamma_{1/2}L_{-1}+\gamma_{-1/2}+
\gamma_{1/2}^2G_{-1/2}b_{-1}-\frac14c_{0}G_{-1/2}\bigg)|\Omega_a\rangle_0,
\\
&Q_3G^{\text{M+L}}_{-1/2}|\Omega_a\rangle_0=-\frac14c_{0}G_{-1/2}|\Omega_a\rangle_0.
\end{aligned}
\end{equation}
The sum of these contributions is
\begin{align}
QG^{\text{M+L}}_{-1/2}|\Omega_a\rangle_0=(\gamma_{1/2}L_{-1}+\gamma_{-1/2}+
\gamma_{1/2}^2G_{-1/2}b_{-1})|\Omega_a\rangle_0.
\label{QWtilde1}
\end{align}
Similarly, the action of the operator $Q$ on the second term in
${\tilde W}_{a}$ is
\begin{equation}
\begin{aligned}
&Q_1b_{-1}\gamma_{1/2}|\Omega_a\rangle_0=\bigg(\gamma_{1/2}L_{-1}+
\frac12\gamma_{-1/2}+\frac34c_0b_{-1}\gamma_{1/2}\bigg)|\Omega_a\rangle_0,
\\
&Q_2b_{-1}\gamma_{1/2}|\Omega_a\rangle_0=\bigg(\gamma_{1/2}^2G_{-1/2}b_{-1}+
\frac12\gamma_{-1/2}-\frac12c_{0}b_{-1}\gamma_{1/2}\bigg)|\Omega_a\rangle_0,
\\
&Q_3b_{-1}\gamma_{1/2}|\Omega_a\rangle_0=
-\frac14 c_{0}b_{-1}\gamma_{1/2}|\Omega_a\rangle_0.
\end{aligned}
\end{equation}
Again summing these contributions, we obtain
\begin{equation}
Qb_{-1}\gamma_{1/2}|\Omega_a\rangle_0=(\gamma_{1/2}L_{-1}+\gamma_{-1/2}+
\gamma_{1/2}^2 G_{-1/2}b_{-1})|\Omega_a\rangle_0.
\end{equation}
This coincides with the the action of the BRST charge on the first
term~\eqref{QWtilde1}. We have thus verified that fields ${\tilde W}_{a}$
form the cohomology class.

\medskip

\appendix{{\bf Appendix C. The basic relation}}

We first consider the action of the holomorphic BRST charge on the discrete
state $O_{m,n}$. We assume that we are dealing with the Verma module in
the Liouville sector (i.e., the degenerate representation is not factored
with respect to the singular vector submodule). From dimensional arguments
and also taking the ghost charge of the operator $Q$ into account, we can
conclude that the most general form of this action is
\begin{equation}
QO_{m,n}=(xG_{-1/2}^{\text{M}}+yG_{-1/2}^{\text{L}}+zG^{\text{g}}_{-1/2})\Phi_{m,n}
D^{\text{L}}_{m,n}V_{m,n}c.
\end{equation}
Indeed, decoupling the singular vector then provides that $O_{m,n}$ is
BRST exact. The explicit calculation gives $x=y=1$ and $z=-1/2$. Combining
with the action of the antiholomorhic $\bar Q$, we obtain basic
relation~\eqref{basic0}. We now consider the ``quasi''-discrete state
\begin{equation}
O_{a}=\bar H_{m,n}H_{m,n}\Phi_{m,n}V_{a}
\end{equation}
for which the parameter $a$ is in the vicinity of the degenerate value or,
more precisely, the difference $\epsilon=a-a_{m,n}$ is small. It is obvious
from the analyticity that
\begin{equation}
\bar QQO_{a}=[\bar J_{m,n}+\epsilon\bar K_{m,n}][J_{m,n}+
\epsilon K_{m,n}]\Phi_{m,n}V_{a},
\label{QQOa}
\end{equation}
where $J_{m,n}$ is defined in~\eqref{Jmn} and $K_{m,n}$ is an operator
built from the super Virasoro generators of all three sectors.
Differentiating~\eqref{QQOa} with respect to the parameter $a$ gives
\begin{align}
\bar QQO'_{a}&=\bar J_{m,n}J_{m,n}\Phi_{m,n}V_{m,n}'
\nonumber
\\
&=\biggl(\bar G_{-1/2}^{\text{M+L}}-\frac12\bar G_{-1/2}^{\text{g}}\biggr)
\biggl(G_{-1/2}^{\text{M+L}}-\frac12G_{-1/2}^{\text{g}}\biggr)
\Phi_{m,n}\bar D^{\text{L}}_{m,n}D^{\text{L}}_{m,n}V'_{m,n}\bar cc
\label{QQO'}
\end{align}
because the term
\begin{equation}
(\bar K_{m,n}J_{m,n}+K_{m,n}\bar J_{m,n})\Phi_{m,n}V_{m,n}
\end{equation}
vanishes as a result of the action of the operators $D_{m,n}^L$
and $\bar D_{m,n}^L$ inside $J_{m,n}$ and $\bar J_{m,n}$ on $V_{m,n}$.
Relation~\eqref{QQO'} combined with higher equations of motion~\eqref{HEM}
results in~\eqref{basic}.

\medskip

\appendix{\bf Appendix D. OPE $O_{1,3}W_a$.}

We separately calculate the contribution of each term in~\eqref{O13} to
the operator product $O_{1,3}(x)W_a(0)$. The first term gives (we write
only the holomorphic part explicitly)
\begin{align}
&\Phi_{13}'(x)V_{13}(x)\Phi_{a-b}(0)V_{a}(0)c(0)\delta(\gamma(0))
\nonumber
\\
&\quad=\bigl(x^{1-a b}C_{+}^{\text{M}}(a-b)\Phi_{a-2b}(0)+
x^{1-b^2}\tilde C_{0}^{\text{M}}(a-b)\Psi_{a-b}(0)+
x^{a b-b^2}C_{-}^{\text{M}}(a-b)\Phi_{a}(0)\bigr)'
\nonumber
\\
&\quad\quad\;\times\bigr(x^{a b}C_{-}^{\text{L}}(a)V_{a-b}(0)+
x^{1+b^2}\tilde C_{0}^{\text{L}}(a)\Lambda_{a-b}(0)+
x^{1-a b+b^2}C_{+}^{\text{L}}(a)V_{a+b}(0)\bigr)C(0)\delta(\gamma(0))
\nonumber
\\
&\quad=(1- a b)C_{+}^{(M)}(a-b)C_{-}(a)W_{a-b}(0)+
(a b-b^2)C_{-}^{(M)}(a-b)C_{+}(a)W_{a+b}(0).
\end{align}
The contribution of the second term in~\eqref{O13} is
\begin{align}
&\Phi_{13}(x)V'_{13}(x)\Phi_{a-b}(0)V_{a}(0)C(0)\delta(\gamma(0))
\nonumber
\\
&\qquad=abC_{+}^{\text{M}}(a-b)C_{-}^{\text{L}}(a)W_{a-b}(0)+
(1-a b+b^2)C_{-}^{\text{M}}(a-b)C_{+}^{\text{L}}(a)W_{a+b}(0).
\end{align}
The third term contributes
\begin{equation}
\Psi_{13}(x)\Lambda_{13}(x)\Phi_{a-b}(0)V_{a}(0)c(0)\delta(\gamma(0))=
\tilde C_{0}^{\text{M}}(a-b)\tilde C_{0}^{\text{L}}(a)W_{a}(0).
\end{equation}
Using the basic operator product in the ghost sector, we obtain
\begin{equation}
\beta(x)\delta(\gamma(0))=\frac{\delta'(\gamma(0))}{x}
\end{equation}
and
\begin{equation}
\gamma(x)\delta'(\gamma(0))\sim
\gamma(0)\delta'(\gamma(0))\sim-\delta(\gamma(0)).
\end{equation}
Hence, the fourth term contributes
\begin{align}
&\Phi_{13}(x)\Phi_{a-b}(0)V_{13}(x)V_{a}(0)\beta(x)
\gamma(x)c(0)\delta(\gamma(0))=\nonumber
\\
&\quad-C_{+}^{\text{M}}(a-b)C_{-}^{\text{L}}(a)W_{a-b}(0)-
C_{-}^{\text{M}}(a-b)C_{+}^{\text{L}}(a)W_{a+b}(0).
\end{align}
The last term contributes
\begin{align}
&\Phi_{13}(x)\Phi_{a-b}(0)V_{13}(x)V_{a}(0)b(x)c(x)c(0)\delta(\gamma(0))=
\nonumber
\\
&\quad C_+^{\text{M}}(a-b)C_-^{\text{L}}(a)W_{a-b}(0)+C_-^{\text{M}}(a-b)C_+^{\text{L}}(a)W_{a+b}(0).
\end{align}
Combining all together and also taking the antiholomorphic part into account,
we obtain~\eqref{O13W}.

\medskip

\appendix{\bf Appendix E. Three-point correlation number in SLG}

Here, we explicitly derive three-point number~\eqref{3point},
\begin{equation}
\langle\langle a_1a_2a_3\rangle\rangle_{\text{SLG}}=
C_{\text{I}}^{\text{SLG}}(a_1,a_2,a_3)+
C_{\text{II}}^{\text{SLG}}(a_1,a_2,a_3),
\end{equation}
where
\begin{equation}
\begin{aligned}
&C_{\text{I}}^{\text{SLG}}(a_1,a_2,a_3)=
C^{\text{L}}(a_1,a_2,a_3)\tilde C^{\text{M}}(a_1-b,a_2-b,a_3-b),
\\
&C_{\text{II}}^{\text{SLG}}(a_1,a_2,a_3)=
\tilde C^{\text{L}}(a_1,a_2,a_3)C^{\text{M}}(a_1-b,a_2-b,a_3-b).
\end{aligned}
\end{equation}
Instead of treating the two sectors separately, we find it instructive to
solve the shift relations for the SLG structure constants. Below, we show
that
\begin{align}
\frac{C_{\text{I,II}}^{\text{SLG}}(a_1+b,a_2,a_3)}
{C_{\text{I,II}}^{\text{SLG}}(a_1-b,a_2,a_3)}={}&
[\pi \mu b^2 \gamma(1/2+b^2/2)]^{-2}(1/2-b a +b^2/2)(1/2-b a -b^2/2)
\nonumber
\\
&{}\times\bigg[\frac{\gamma(a b +b^2/2+1/2)}
{\gamma(a b-3b^2/2+1/2)}\bigg]^{1/2}.
\label{3p0}
\end{align}
These shift relations give
\begin{equation}
\langle\langle a_1a_2a_3\rangle\rangle_{\text{SLG}}=
\text{const}\prod_{i=1}^3N(a_i),
\end{equation}
where $N(a)$ is defined in Eq.~\eqref{N}. To define the constant, we take
$a_i=b$. The corresponding structure constants in the matter sector are
$C^{\text{M}}(0,0,0)=1$ and $\tilde C^{\text{M}}(0,0,0)=0$. Also having
in mind that $C^{\text{L}}(b,b,b)$ can be defined from~\eqref{C3} and is
expressed just in terms of the gamma functions, we conclude that
$\text{const}=\Omega(b)$ (see Eq.~\eqref{Omega}). The shift relations for
SLG structure constants~\eqref{3p0} are combined from the corresponding
shift relations in the Liouville and the matter sectors,
\begin{equation}
\frac{C_{\text{I}}^{\text{SLG}}(a_1+b,a_2,a_3)}
{C_{\text{I}}^{\text{SLG}}(a_1-b,a_2,a_3)}=
\frac{C^{\text{L}}(a_1+b,a_2,a_3)}{C^{\text{L}}(a_1-b,a_2,a_3)}
\frac{\tilde C^{\text{M}}(a_1,a_2-b,a_3-b)}
{\tilde C^{\text{M}}(a_1-2b,a_2-b,a_3-b)},
\end{equation}
which follows from the bootstrap relations and the monodromy properties
of the differential equation corresponding to the decoupling of the
singular vector. For the sake of completeness, we recapitulate the two main
formulas concerning the Liouville sector in~\cite{BBNZ,VB1}. The first is
\begin{align}
\frac{C_-^{\text{L}}(a_1)C^{\text{L}}_{a_{1}+b,a_{2},a_{3}}}
{C_+^{\text{L}}(a_1)C^{\text{L}}_{a_1-b,a_2,a_3}}={}&
-\frac{\gamma(ba_1)\gamma(ba_1-b^2)\gamma^2(1/2-b^2/2+ba_1)}
{(1/2-ba_1+b^2/2)^2}
\nonumber
\\
&{}\times\frac{\gamma(1/2+ba_{2+3-1}/2)\gamma(ba_{2+3-1}/2-b^2/2)}
{\gamma(1/2+ba_{1+3-2}/2)\gamma(ba_{1+3-2}/2-b^2/2)}
\nonumber
\\
&{}\times\frac{\gamma(3/2-ba_{1+2+3}/2+b^2)\gamma(1-ba_{1+2+3}/2+b^2/2)}
{\gamma(1/2+ba_{1+2-3}/2)\gamma(ba_{1+2-3}/2-b^2/2)},
\label{CpCp}
\end{align}
and the second is
\begin{align}
\frac{C_-^{\text{L}}(a_1)\tilde C^{\text{L}}(a_1+b,a_2,a_3)}
{C_+^{\text{L}}(a_1)\tilde C^{\text{L}}(a_1-b,a_2,a_3)}={}&
\frac{\gamma(ba_1)\gamma(ba_1-b^2)}
{\gamma(ba_{1+2-3}/2)\gamma(ba_{1-2+3}/2)}
\nonumber
\\
&{}\times\frac{\gamma(a_1b-b^2/2+1/2)\gamma(a_1b-b^2/2-1/2)}
{\gamma(ba_{1+2+3}/2-b^2)\gamma\bigl((1+ba_{1-2+3}-b^2)/2\bigr)}
\nonumber
\\
&{}\times\frac{\gamma(ba_{-1+2+3}/2)
\gamma\bigl((1+ba_{-1+2+3}-b^2)/2\bigr)}
{\gamma\bigl((-1+b a_{1+2+3}-b^2)/2\bigr)
\gamma\bigl((1+b a_{1+2-3}-b^2)/2\bigr)}
\nonumber
\\
={}&S(a_1,a_2,a_3).
\label{rel2}
\end{align}
Taking~\eqref{C+-0Liouv} into account, we derive
\begin{align}
\frac{C^{\text{L}}(a_1+b,a_2,a_3)}
{C^{\text{L}}(a_1-b,a_2,a_3)}={}&
\biggl[\pi\mu b^2\gamma\biggl(\frac{Qb}2\biggr)\biggr]^{-2}
\frac{\gamma(ba_1)\gamma(ba_1-b^2)}
{\gamma(1/2+ba_{1+3-2}/2)}
\nonumber
\\
&{}\times\frac{\gamma(a_1b+b^2/2+1/2)\gamma(a_1b-b^2/2+1/2)}
{\gamma(ba_{1+3-2}/2-b^2/2)}
\nonumber
\\
&{}\times\frac{\gamma(1-ba_{1+2+3}/2+b^2/2)\gamma(3/2-ba_{1+2+3}/2+b^2)}
{\gamma(1/2+ba_{1+2-3}/2)}
\nonumber
\\
&{}\times\frac{\gamma(1/2+ba_{2+3-1}/2)\gamma(ba_{2+3-1}/2-b^2/2)}
{\gamma(ba_{1+2-3}/2-b^2/2)}
\label{3p1}
\end{align}
from~\eqref{CpCp}.
The shift relations for the structure constants in the matter sector
can be derived in two steps starting from those in the Liouville sector.
First, we change the normalization in accordance with the standard
requirement $\langle\Phi_a\Phi_a\rangle=1$ in the matter sector. In this
normalization, the special structure constants coincide with the
three-point functions (to distinguish these structure constants from the
original ones, we do not label them with the superscript L):
\begin{equation}
\begin{aligned}
&C_-(a)=C(-b,a,a+b),
\\
&C_+(a)=C(-b,a,a-b).
\end{aligned}
\end{equation}
Setting $a_2=a_1$ and $a_3=-b$ in~\eqref{CpCp}, we obtain
\begin{equation}
\bigg[\frac{C_-(a_1)}{C_+(a_1)}\bigg]^2=
\frac{\gamma(a_1b -b^2/2+1/2)\gamma(-a_1b+3b^2/2+3/2)}
{\gamma(-a_1b +b^2/2+3/2)\gamma(a_1b+b^2/2+1/2)}=M(a_1).
\end{equation}
Combining this relation with~\eqref{rel2}, we derive
\begin{equation}
\frac{\tilde C(a_1+b,a_2,a_3)}{\tilde C(a_1-b,a_2,a_3)}=
S(a_1,a_2,a_3)M^{-1/2}(a_1).
\end{equation}
The second step now is to replace $a_i\to i(a_i-b)$ and $b\to-ib$,
\begin{align}
\frac{\tilde C^{\text{M}}(a_1,a_2-b,a_3-b)}
{\tilde C^{\text{M}}(a_1-2b,a_2-b,a_3-b)}={}&
\frac{M^{1/2}(i(a_1-b);-i b)}{S(i(a_1-b),i(a_2-b),i(a_3-b);-i b)}
\nonumber
\\
={}&\frac{\gamma(a_1 b -b^2/2+1/2)\gamma(-a_1 b-b^2/2+3/2)}
{\gamma(-a_1 b +b^2/2+3/2)\gamma(a_1 b-3b^2/2+1/2)}
[\gamma(ba_1)]^{-1/2}
\nonumber
\\
&{}\times[\gamma(ba_1-b^2)\gamma(a_1b+b^2/2+1/2)
\gamma(a_1b-b^2/2+1/2)]^{-1/2}
\nonumber
\\
&{}\times\frac{\gamma(1/2+ba_{1+3-2}/2)\gamma(ba_{1+3-2}/2-b^2/2)}
{\gamma(1-ba_{1+2+3}/2+b^2/2)\gamma(3/2-ba_{1+2+3}/2+b^2)}
\nonumber
\\
&{}\times\frac{\gamma(1/2+ba_{1+2-3}/2)\gamma(ba_{1+2-3}/2-b^2/2)}
{\gamma(1/2+ba_{2+3-1}/2)\gamma(ba_{2+3-1}/2-b^2/2)}.
\label{3p2}
\end{align}
We note that the dependence on the gamma functions containing the different
combinations of $a_i$ in~\eqref{3p1} and~\eqref{3p2} exactly cancel, and we
obtain~\eqref{3p0}.

\end{document}